\pdfoutput=1
\documentclass[conference]{IEEEtran}
\usepackage{cite}
\usepackage{amsmath,amssymb,amsfonts}
\usepackage{algorithmic}
\usepackage{graphicx}
\usepackage{textcomp}
\usepackage{hyperref}
\usepackage{subcaption}
\usepackage{stfloats}
\usepackage{xurl}
\usepackage{xcolor}
\hypersetup{pdfauthor={}, pdftitle={Anonymous Submission}, pdfsubject={}, pdfkeywords={}}
\renewcommand{\figurename}{Figure}

\def\BibTeX{{\rm B\kern-.05em{\sc i\kern-.025em b}\kern-.08em
    T\kern-.1667em\lower.7ex\hbox{E}\kern-.125emX}}

\begin{document}

\title{Green Light: Carbon-Aware Design of Photonic Transformer Accelerators at the Edge}

 \author{\IEEEauthorblockN{Bharat Bhatta, Ishan Thakkar}
 \IEEEauthorblockA{Department of Electrical and Computer Engineering, University of Kentucky, Lexington, KY 40506, USA\\
 bharat.bhatta@uky.edu, igthakkar@uky.edu}
 }

\maketitle

\begin{abstract}
Migrating transformer inference to the edge requires sustainability considerations that encompass both operation and manufacturing, rather than operational energy alone. Silicon photonics enables high-bandwidth, low-energy matrix multiplication; however, existing microring-resonator-based Dot Product Unit (DPU) organizations have been tailored to convolutional computations and evaluated using analog signaling without accounting for embodied carbon footprints. We adapt three DPU organizations (AMW, MAW, and MWA) to transformer inference workloads and extend each organization to stochastic signaling, yielding six organizations in total. We then calculate their embodied and operational carbon per inference. Across five transformer models and four foundry grids, the photonic system achieves improvements of four, four, and six orders of magnitude in Energy--Delay, Carbon--Delay, and Carbon--Energy Product, respectively, relative to CPUs, while also being compared against H100 SXM5 GPU and TPU v5p baseline systems. Analog-to-stochastic conversion increases die area by 4.3--4.6$\times$ and embodied carbon by 4.4--4.8$\times$, while reducing operational energy by approximately 91\% and causing embodied carbon to equal or exceed operational carbon per inference. Importantly, the DPU organization and signaling scheme, rather than the foundry grid, are the key determinants of carbon impact. Consequently, the fastest organization (MWA) is not necessarily the most carbon-efficient, whereas MAW-Stochastic offers a favorable balance between sustainability and performance.
\end{abstract}


\section{Introduction}
Transformer inference is shifting toward the edge, where per-query latency, energy, and carbon are primary constraints \cite{most_used}, and their aggregate impacts increase with the intensity of model use. A single ChatGPT prompt emits several grams of CO$_2$e, substantially more than a web search \cite{r2}; across billions of daily queries, these emissions accumulate rapidly. Meanwhile, model sizes have increased from 1.5 billion parameters in GPT-2 to 175 billion in GPT-3 \cite{survey}, and the ICT sector now accounts for an estimated 4.1\% of global greenhouse gas emissions \cite{r3}. Because the majority of this footprint stems from serving hardware, edge accelerators must be evaluated based on their combined manufacturing and operational carbon, rather than inference energy alone.

\begin{figure}[htbp]
    \centering
    \includegraphics[width=\columnwidth]{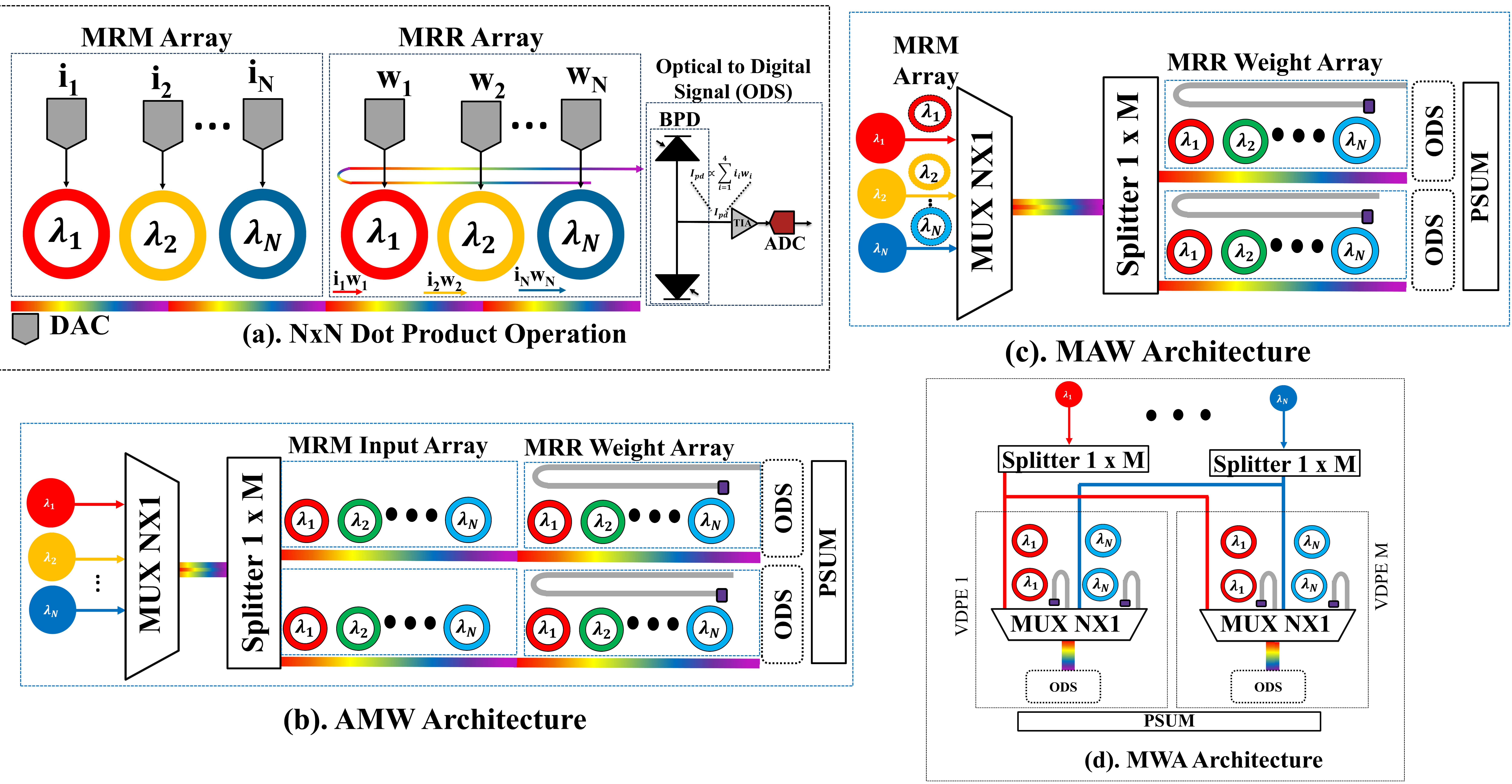}
    \captionsetup{font=small}
    \caption{Microring-based photonic DPU operation and the three analog DPU organizations. (a) An MRM array encodes inputs and an MRR array applies weights, each on a distinct wavelength, with products summed at a balanced photodetector (BPD) and digitized by an ADC. (b) AMW, (c) MAW, and (d) MWA, each built from Vector Dot Product Elements (VDPEs) with optical-to-digital signal (ODS) conversion and partial-sum reduction (PSUM).}
    \label{Architectures}
\end{figure}

Silicon photonic accelerators have shown promise as a substrate for edge transformer inference because they perform computation using light, thereby providing high bandwidth and low latency at substantially lower energy per operation, particularly for matrix multiplication \cite{r4,r5,r6,deap_CNN,holylight}. Each chip integrates multiple MRR-based Dot Product Units (DPUs). A DPU contains $M$ vector dot product elements (VDPEs), each of which spans $N$ wavelengths to compute an $N$-length dot product, with one value per wavelength. Within a DPU, the wavelengths pass through Splitting (S), Aggregation (A), Modulation (M), and Weighting (W) stages. The ordering of these stages defines three organizations: AMW, MAW, and MWA \cite{comparative}. The AMW organization includes Crosslight, DEAPCNN, Robin, and RAMM; the MASW organization is adopted by Holylight, Yang, Al-Qadasi, PCNNA, and RMAM; and the SMWA organization is adopted by Hitless, ADEPT, Albireo, and ASTRA \cite{comparative,astra}. This ordering determines the accumulation of optical loss, inter-channel crosstalk, and the $(N,M)$ dimensions that fit within a given area and power budget \cite{tait2016microring,bahadori2016comprehensive,padmaraju2014intermodulation}, resulting in a trade-off among signal quality, scalability, and area (Figure~\ref{Architectures}).

These organizations, however, were designed for convolutional neural networks \cite{comparative,he2016deep}, whose operational dataflow and dependencies differ substantially from the sequence-dependent and head-dependent attention mechanisms that characterize transformers and scale quadratically with sequence length \cite{ashish2017attention,tay2022efficient}. Signal representation is another important consideration. Analog systems represent information using multi-level optical amplitudes, requiring expensive, power-hungry DACs and a wide dynamic range, whereas stochastic systems use pulses that are multiplied by optical AND gates \cite{sconna,astra}.

We adapt all three DPU organizations to transformer inference, extend each organization to stochastic signaling, and compare the resulting six configurations against CPU, GPU, and TPU baselines. This paper makes the following three contributions:
\begin{itemize}
    \item We adapt all three MRR-based DPU organizations (AMW, MAW, and MWA) to transformer inference, derive their scalability-limited $(N,M)$ dimensions, and develop a dataflow with fused Q/K/V projection and comb-switch attention packing.
    
    \item We extend each organization to stochastic signaling, yielding six configurations across two signaling paradigms, and model the die area and embodied carbon of each configuration.
    
    \item We report the first evaluation of both operational and embodied carbon for photonic transformer accelerators. This evaluation incorporates 2.5D interposer integration overheads and compares the designs using carbon-aware metrics across four foundry grids.
\end{itemize}

Our analysis yields a non-obvious result: at per-inference granularity, embodied carbon exceeds operational carbon. Consequently, the fastest organization (MWA) is not the most carbon-efficient; instead, MAW-Stochastic achieves the highest carbon efficiency. Thus, operational efficiency and carbon efficiency diverge at the edge.

\section{Background}
\subsection{Photonic GEMM Accelerators and DPU Organizations}
Transformer inference is dominated by General matrix-matrix multiplication (GEMM). Its regular structure maps naturally onto wavelength-parallel photonic hardware, motivating a range of MRR-based accelerators \cite{r4,r5,r6,deap_CNN,holylight}.  
Each architectures introduce in Section I implement the same GEMM using varying ordering. Each architecture is not equivalent to another because each has specific optical loss and wavelength channel crosstalk losses that determine their DPUs density, and thereby $(N, M)$ pair that they can support. Earlier research studied the performance of these architectures assuming analog signals. So it is unknown how they compare once embodied carbon and stochastic operation are considered. We address this gap.

\subsection{Analog vs. Stochastic Optical Computing}


Analog accelerators require input DACs and a wide optical dynamic range to distinguish signal levels amid noise and crosstalk, increasing energy and peripheral area \cite{sunny2021crosslight,deap_CNN,holylight,sunny2021robin}. In contrast, stochastic accelerators encode values as ON/OFF bitstreams and multiply using optical AND gates, requiring only binary discrimination \cite{ren2017sc,sconna,astra}.

\subsection{Architectural Carbon Modeling}
 Carbon modeling has matured from one-off fabrication assessments into reusable architectural frameworks. 
After works which quantify the energy, water, and emissions cost of logic-node fabrication \cite{bardon2020dtco}, ACT introduces component-level estimates with sustainability-aware metrics \cite{act}, and later frameworks add chiplet packaging \cite{eco} and probabilistic lifecycle estimation across a decade of processors, which found aggregate emissions trending upwards in the era of AI-driven production despite per-chip improvements \cite{hu2025carbonset}, for a deployed photonic accelerator there are two threads of work which miss out: electronic chiplet models captures die-to-die interfaces, bonding, and yield for 2.5D and 3D integration, but assume all-electronic dies \cite{feng2022chiplet,zhao20243d}, whereas EPiCarbon models electro-photonic dies but not their package-level integration \cite{epicarbon}. We combine them and incorporate 2.5D interposer integration overheads into embodied-carbon modeling of photonic-electronic chiplet systems. We use it for the first per-inference, embodied and operational comparison of photonic transformer accelerators across organizations and signaling modes.

\section{From Analog to Stochastic: Six Photonic Transformer Accelerators}
\label{sec:arch}
By pairing each of the three DPU organizations (AMW, MAW, and MWA) with analog and stochastic signaling, we successfully construct six photonic accelerator configurations.The two modes of signaling share identical stage arrangement, and vary only in the way that the operands are modulated and multiplied to ensure a fair comparison. Thus, the difference in observation made through the signaling is independent of the structure. All six topologies are evaluated at the INT8 precision level. Firstly, the analog baselines and their scalability limitations are discussed. Secondly, the stochastic variations are introduced, followed by the iso-precision assessment for both. Thirdly, the non-linear support and the dataflow mapping are introduced..

\subsection{Analog Baselines and Their Scalability Limits}
\label{sec:analog}
The analog baselines execute transformer GEMM operations entirely in the optical domain using MRR-based DPUs \cite{comparative}. DAC arrays convert digital operands into multi-level optical amplitudes that drive microring modulators (MRMs), and values are combined by wavelength-division multiplexing. The weighted optical signals are detected at balanced photodetectors, converted to voltage by a transimpedance amplifier (TIA), digitized by an ADC, and accumulated in a partial-sum reduction network (PSUM) \cite{deap_CNN,holylight,comparative,recongigure}. Digital conversion is unavoidable due to the lack of native support for Softmax and activation layers in the optical domain \cite{astra}.

The scale of each organization is bounded by the optical link budget. Each chip contains multiple DPUs, whose number we call the DPU count; each DPU holds $M$ vector dot product elements (VDPEs), and each VDPE spans $N$ wavelengths and therefore computes an $N$-length dot product. 
As the number of VDPEs ($N$) and their amount per DPU ($M$) in the system increases, the overall insertion loss and inter channel crosstalk lower the optical power launched into each photodetection device and the design becomes non-feasible when the level reaches below the photosensitivity.
We therefore determine the largest feasible $(N, M)$ at a 4-bit per-slice DAC resolution and a 1~GS/s data rate. Substituting the parameters of Table~\ref{tab:abbreviations_parameters} into Eq.~\eqref{eq:bit} and Eq.~\eqref{eq:beta_cal} and solving simultaneously yields the required photodetector sensitivity $P_{PD-opt}$, which the link budget of Eq.~\eqref{eq:link_budget} then converts, for a fixed $N$, into the maximum $M$ that keeps accumulated loss and crosstalk within budget \cite{scaling}. The analog block of Table~\ref{tab:m_values} lists the resulting pairs. 
We adopt $N=128$ for all three organizations due to the fact that it maximizes the aggregate dot-product parallelism $N \times M$ within the sensitivity budget, and thus the number of concurrent MACs per DPU.

{\footnotesize
\begin{equation}
\label{eq:bit}
B_{Res} = \frac{1}{6.02} \left[ 20\log_{10}\!\left(\frac{R \times P_{PD-opt}}{\beta\sqrt{DR/\sqrt{2}}}\right) - 1.76 \right]
\end{equation}
}

{\footnotesize
\begin{equation}
\label{eq:beta_cal}
\begin{split}
\beta =& \sqrt{2q(RP_{PD-opt} + I_d) + \frac{4kT}{R_L} + R^2 P_{PD-opt}^2 RIN} + \sqrt{2qI_d + \frac{4kT}{R_L}}
\end{split}
\end{equation}
}

{\footnotesize
\begin{equation}
\label{eq:link_budget}
\begin{split}
P_{PD-opt}(\mathrm{dBm}) = P_{Laser} - IL_{SMF} - IL_{EC} 
\quad - IL_{WG} \times N \times d_{OSM} \\
\quad - IL_{OSM} - (N-1)\,OBL_{OSM} 
\quad - EL_{Splitter} \times \log_2(M) \\
\quad - IL_{MRR} - (N-1)\,OBL_{MRR} 
\quad - IL_{penalty} - 10\log_{10}(M)
\end{split}
\end{equation}
}

\begin{table}[htbp]
\captionsetup{font=small}
\caption{Definition and values of parameters (from \cite{scaling,comparative,recongigure,astra,sconna}) used in Eq.~\eqref{eq:bit} and \eqref{eq:link_budget} for the scalability analysis of both analog and stochastic accelerators. $BR$ is stochastic bitrate and takes the role of $DR$ in Eq.~\eqref{eq:bit}; the analog analysis uses $DR = 1$~GS/s.}
\label{tab:abbreviations_parameters}
\centering
\begin{tabular}{|c|c|c|c|}
\hline
\textbf{Parameter} & \textbf{Value} & \textbf{Parameter} & \textbf{Value} \\
\hline
$P_{Laser}$ & 10 dBm & $EL_{Splitter}$ (dB) & 0.01 \\
\hline
$R_{PD}$ & 1.2 A/W & $IL_{OSM}$ (dB) & 4 \\
\hline
$R_L$ & 50 $\Omega$ & $OBL_{OSM}$ (dB) & 0.01 \\
\hline
$I_d$ & 35 nA & $IL_{MRR}$ (dB) & 0.01 \\
\hline
$T$ & 300 K & $OBL_{MRR}$ (dB) & 0.01 \\ 
\hline
$BR$ & 30 GS/s & $IL_{penalty}$ (dB) & 5.3 \\
\hline
$RIN$ & -140 dB/Hz & $IL_{SMF}$ (dB) & 0 \\
\hline
$\eta_{WPE}$ & 0.1 & $IL_{EC}$ (dB) & 1.6 \\
\hline
$d_{OSM}$ & 20 $\mu$m & $IL_{WG}$ (dB/mm) & 0.3 \\
\hline
\end{tabular}
\end{table}

\vspace{10pt}

\begin{table}[htbp]
\centering
\captionsetup{font=small}
\caption{Scalability-limited VDPE size ($N$) and VDPE count ($M$) for all six configurations. Analog: 4-bit per-slice DAC at $DR = 1$~GS/s; stochastic: 1-bit per-cycle stream at $DR = 30$~GS/s. Both reach INT8 precision\cite{astra, sconna} (Section~\ref{sec:precision}) via bit-slicing and 256-bit bitstream encoding respectively. We adopt $N=128$.}
\label{tab:m_values}
\begin{tabular}{|l|c|c|c|c|c|}
\hline
 & \textbf{N=128} & \textbf{N=64} & \textbf{N=32} & \textbf{N=16} & \textbf{N=8} \\
\hline
\multicolumn{6}{|c|}{\textbf{Analog (4-bit per slice, 1~GS/s)}} \\
\hline
AMW & 17 & 27 & 35 & 39 & 42 \\
MAW & 35 & 44 & 50 & 53 & 54 \\
MWA & 93 & 102 & 106 & 109 & 110 \\
\hline
\multicolumn{6}{|c|}{\textbf{Stochastic (256-bit stream, 30~GS/s)}} \\
\hline
AMW & 88 & 141 & 178 & 201 & 213 \\
MAW & 178 & 225 & 253 & 269 & 277 \\
MWA & 472 & 516 & 539 & 551 & 557 \\
\hline
\end{tabular}
\end{table}

\subsection{Stochastic Variants: Trading Amplitude for Bitstreams}
\label{sec:stochastic}
In Stochastic signaling operands are encoded as bit- streams and multiplied with simple logic rather than resolved as precise analog levels. We converted each analog architecture to a stochastic variant through two changes. First, the DAC driven modulation stage is replaced by Optical Stochastic Multipliers (OSMs) that emit binary ON/OFF pulse streams. Second, multiplication is performed explicitly by Optical AND Gates (OAGs) rather than implicitly from the analog signal combination. Together, these changes eliminated the DAC stage and the need for wide optical dynamic range: lookup tables mapped binary operands to stochastic bitstreams, serializers fed the OSMs, and an integrator replaced the TIA to accumulate the AND-gate outputs within the bitstream duration \cite{Integrator_simphony}. We apply this modification uniformly across all three organizations and preserve the stage ordering, so that any difference between an analog design and its stochastic counterpart is attributable to signaling alone.

This shift in signaling directly loosens the sensitivity constraint that bounded the analog designs. Because a stochastic receiver need only distinguish photon-present from photon-absent states rather than resolve multiple amplitude levels, its sensitivity requirement relaxes and the dark-current noise term $\sqrt{2qI_d + 4kT/R_L}$ in Eq.~\eqref{eq:beta_cal} drops out. Re-applying the procedure of Section~\ref{sec:analog} with 1-bit resolution and a 30~GS/s data rate yields the stochastic block of Table~\ref{tab:m_values}.The net effect is large. At N = 128, M increases from 17 to 88 for AMW, from 35 to 178 for MAW, and from 93 to 472 for MWA. Which is around fivefold increase in per-DPU parallelism, which later makes up for the longer bitstream latency. The resulting AMW-Stoch design corresponds to the SCONNA accelerator \cite{sconna}, and the MWA-Stoch design corresponds to the ASTRA accelerator \cite{astra}.

\subsection{Iso-Precision INT8 Evaluation}
\label{sec:precision}
Each INT8 operand is encoded into a stochastic bitstream of 256 bits using OSMs. The multiplication between operands is performed by the optical AND gates over the whole bitstream, followed by integration of the output. Analog schemes employ 4-bit DACs; therefore, the precision is increased to 8-bit using bit-slicing. Each 8-bit operand is divided into two 4-bit slices, and each slice is calculated separately through optical processing, later they are aggregated. We take into account the complexity of both optical processes, thus making sure that the analog latencies, energy, and areas correspond to the full 8-bit computation process. Thus, we make sure that all six schemes can be compared in terms of workloads.

\subsection{Non-Linear Operations on a Shared Digital Path}
\label{sec:nonlinear}
Operations that can’t be computed by direct light are done via a common digital pipeline, where the previously mentioned stochastic photonic accelerators \cite{astra}. In softmax, max-subtraction is implemented using the comparator, along with adders and look-up tables for exponential and reciprocal. GELU uses a look-up table and adder. The input to LayerNorm is fed to an adder tree and then used in the inverse square root operation using the look-up table. As all organizations use the same path, it introduces the same overhead to all of them. 

\subsection{Transformer Dataflow and Layer Mapping}
\label{sec:dataflow}
Mapping each transformer layer onto the DPU array as a pipeline of stages is where we exploit the transformer-specific structure. For a given configuration $(N, M, \text{DPU count})$ and model, each layer runs as an ordered sequence: Q/K/V projection, $\text{QK}^\top$, Softmax, attention, output projection, two LayerNorm stages, and a two-layer feed forward block with GELU. We decomposed each GEMM stage's energy into input conversion (DAC or serializer), non-input processing (ADC and detector), and partial-sum reduction, which shows two optimizations. First, the fused Q/K/V projection, which transforms the shared input activation only once, but charges non-input and partial sum energy three times. Second, for attention with per-head dimension $K\le 64$, comb-switch packing \cite{recongigure} enables a single N = 128 VDPE to execute two 64-wide dot products in parallel, reducing VDPE count for the $\text{QK}^\top$ and attention-value stages by a factor of 2. Per-layer results are multiplied by the number of layers, which gives a combined term of a weighted static-energy term, and energy over all layers. Here, operand reuse and comb-switch packing concentrate their savings into the attention operations.

\section{Carbon Footprint Framework}
\subsection{Carbon Modeling Methodology}
To assess sustainability at the edge for each inference, we need both manufacturing and operational emissions. We therefore define the total carbon per inference as

{\footnotesize
\begin{equation}
\label{eq:cf_inf}
    CF_{Inference} = CF_{operational} + \frac{T_{exec}}{LT}\,CF_{embodied}
\end{equation}
}

where $T_{exec}$ is the per-inference execution time and $LT$ is the hardware lifetime, so embodied carbon is amortized over every inference the hardware performs in its lifetime. We assume Lt as 3-year hardware lifetime, consistent with typical hardware refresh cycles\cite{act}. Operational carbon is the energy cost of that inference,

{\footnotesize
\begin{equation}
\label{eq:cf_op}
    CF_{operational} = E_{total} \times CI_{grid}
\end{equation}
}

where $E_{total}$ is the per-inference energy and $CI_{grid}$ is the operational-grid carbon intensity in $gCO_2/\mathrm{kWh}$.

Embodied carbon comprises the CMOS die, the photonic die, and packaging,

{\footnotesize
\begin{equation}
\label{eq:cf_emb_split}
    CF_{embodied}=CF_{CMOS}+CF_{Photonics}+CF_{Packaging}
\end{equation}
}

which we compute with the EPiCarbon\cite{epicarbon}: die-level carbon follows ACT \cite{act} from die area, process node, and defect density, and packaging carbon follows ECO-CHIP \cite{eco}. Each die contributes

\begin{figure}[htbp]
    \centering
    \includegraphics[width=\columnwidth]{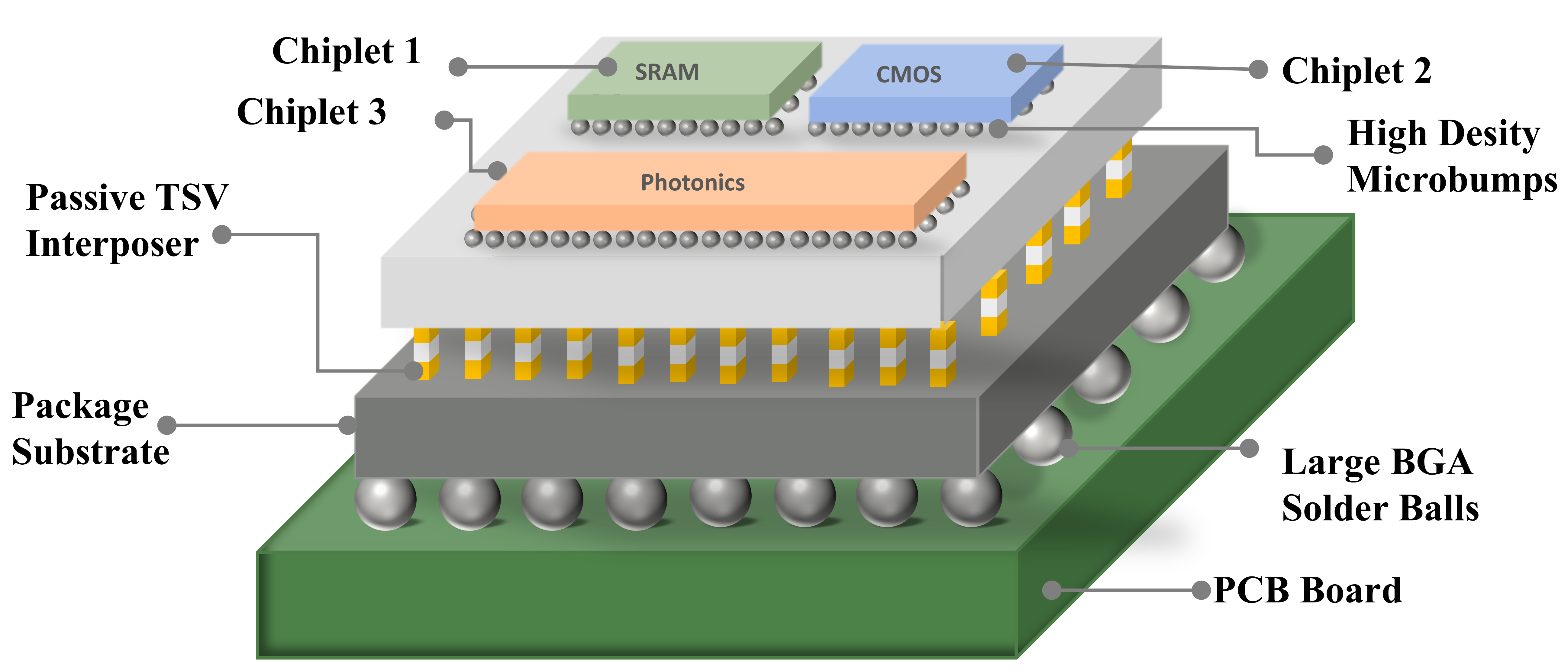}
    \captionsetup{font=small}
    \caption{2.5D packaging architecture. CMOS and photonic chiplets are co-integrated on a passive TSV interposer with SRAM, connected to the package substrate through high-density microbumps and to the PCB board via BGA solder balls.}
    \label{fig:packaging}
\end{figure}

{\footnotesize
\begin{equation}
\label{eq:embo}
    CF_{die} = \frac{\mathrm{Area}}{\mathrm{Yield}} \times (\mathrm{EPA} \times CI_{fab} + \mathrm{GPA} + \mathrm{MPA}) + C_{package}
\end{equation}
}

where Area is the die area of the organization, Yield follows a Poisson defect-density model, EPA, GPA, and MPA are the energy, greenhouse-gas, and materials cost per unit area, and $CI_{fab}$ is the manufacturing-grid carbon intensity. 
We apply EPiCarbon’s photonic-process values of EPA, GPA, and MPA to the photonic die and the corresponding CMOS-node values to the electronic die. Separating $CI_{fab}$ from $CI_{grid}$enables a chip to be fabricated and deployed in different grids, which isolates the fabrication location as a variable in our results.
The paper presents a thorough analysis of the die costs, but beyond the die terms we also incorporate 2.5D interposer integration overheads: as the CMOS and photonic chiplets are co integrated on a passive TSV interposer, as shown in Figure ~\ref{fig:packaging}, which is not modeled by EPiCarbon.

\subsection{Evaluation Metrics}
We assess each configuration using one carbon-independent metric and two carbon-aware metrics from ACT \cite{act}. The Energy-Delay Product, $EDP=E_{total} \times T_{exec}$, balances energy and latency; the Carbon-Delay Product, $CDP=CF_{Inference} \times T_{exec}$, balances carbon and latency; and the Carbon-Energy Product, $CEP=CF_{Inference} \times E_{total}$, balances carbon and energy. Together, these metrics reveal when performance-optimal and carbon-optimal designs diverge.

\section{Experimental Setup}

\subsection{Simulation Environment}
A custom Python simulator evaluates all six configurations, computing per-pipeline-stage latency and energy for transformer inference under the link-budget derivations of Section~III. Energy accumulates dynamic and static components across MRR tuning, serialization (stochastic mode only), photodetection, and ADC readout. In a post-processing step, we apply the EPiCarbon methodology to the die area and process node from the simulator's DPU count and area decomposition to obtain embodied carbon. Operational carbon and all three metrics then follow directly from the per-inference latency and energy the simulator reports.



\subsection{Workloads}
The five different model types Table~\ref{tab:transformer_models} that we evaluate correspond to a range of architectural sizes and domains, including: Transformer-base (machine translation), BERT-base and ALBERT-base (encoder only language models), ViT-base (vision), and OPT-350 (decoder only and generative). All models are evaluated with batch size one: the more latency sensitive real world scenario.

\vspace{7pt}

\begin{table}[htbp]
\centering
\captionsetup{font=small}
\caption{Transformer model configurations. $L$ is sequence length; $\text{d}_{\text{model}}$ and $\text{d}_{\text{ff}}$ are the model and feed-forward dimensions.}
\label{tab:transformer_models}
\begin{tabular}{lllllll}
\hline
Model & Params & Layers & L & Heads & $\text{d}_{\text{model}}$ & $\text{d}_{\text{ff}}$ \\
\hline
Transformer-base & 52M & 2 & 128 & 8 & 512 & 2048 \\
BERT-base & 108M & 12 & 128 & 12 & 768 & 3072 \\
ALBERT-base & 12M & 12 & 128 & 12 & 768 & 3072 \\
ViT-base & 86M & 12 & 256 & 12 & 768 & 3072 \\
OPT-350 & 350M & 12 & 2048 & 12 & 768 & 3072 \\
\hline
\end{tabular}
\end{table}



\subsection{Grid Carbon Intensity}
As photonic and silicon dies may be fabricated elsewhere than the location at which they are deployed, we treat manufacturing and operational grids as separate entities. Table ~\ref{tab:grid_ci} lists the four foundry grids, spanning a 1.76$\times$ intensity range, and the fixed operational grid. Thus, fabrication location dictates per-configuration embodied carbon, while leaving operational carbon constant, isolating location as a variable in our results.

\vspace{7pt}
\begin{table}[htbp]
\centering
\captionsetup{font=small}
\caption{Carbon intensity of the four foundry manufacturing grids and the fixed operational grid. AZNM and RFCW are the eGRID subregions for Arizona and the US operational average, respectively.}
\label{tab:grid_ci}
\begin{tabular}{|l|c|c|}
\hline
\textbf{Grid} & \textbf{Type} & \textbf{CI ($gCO_2$/kWh)} \\
\hline
Taiwan & Manufacturing & 563.0 \cite{TaiwanEl28:online} \\
\hline
Singapore & Manufacturing & 402.0 \cite{singa} \\
\hline
Israel & Manufacturing & 396.6 \cite{isreal} \\
\hline
Arizona (AZNM) & Manufacturing & 319.2 \cite{usa} \\
\hline
USA (RFCW) & Operational & 413.4 \cite{usa} \\
\hline
\end{tabular}
\end{table}

\subsection{Hardware Parameters}
Table~\ref{tab:accelerator_peripherals} gives the power, latency, and area of every electronic and photonic component in the simulator using published silicon. In analog signaling configurations use a 4-bit DAC, a 1~GS/s ADC, and the analog photodetector, whereas stochastic configurations replace these with a serializer, an integrator, and the digital photodetector. Reduction network, eDRAM, router, IO interface, and the non-linear units (comparator, adder, and LUT) are shared across all six configurations and contribute equally to their absolute cost.

\vspace{7pt}
\begin{table}[htbp]
\centering
\captionsetup{font=small}
\caption{Power, latency, and area of accelerator peripheral and compute components, taken from \cite{comparative,recongigure,astra,sconna}}
\label{tab:accelerator_peripherals}
\resizebox{\columnwidth}{!}{%
\begin{tabular}{|c|c|c|c|}
\hline
 & \textbf{Power (mW)} & \textbf{Latency (ns)} & \textbf{Area} \\
\hline
\textbf{Reduction Network} & 0.050 & 3.125 & 1.66E-01 ($mm^2$)\\
\hline
\textbf{eDRAM} & 41.1 & 1.56 & 3.00E-5 ($mm^2$)\\
\hline
\textbf{Router} & 42 & 2 cycles & 0.151 ($mm^2$)\\
\hline
\textbf{Activation Unit} & 0.52 & 0.78 & 6.00E-04 ($mm^2$)\\
\hline
\textbf{IO Interface} & 140.18 & 0.78 & 2.44E-02 ($mm^2$)\\
\hline
\textbf{Photodetector Digital \cite{pd,pd2}} & 1.1 & 0.037 & 10$\times$20 ($\mu m^2$) \\
\hline
\textbf{Photodetector Analog \cite{pd,pd2}} & 1.1 & 0.294 & 180$\times$66 ($\mu m^2$)\\
\hline
\textbf{TIA \cite{TIA}} & 3 & - & 50 ($\mu m^2$) \\
\hline
\textbf{Integrator \cite{Integrator_simphony}} & 0.3 & - & 560 ($\mu m^2$) \\
\hline
\textbf{Comparator \cite{astra}} & 0.6237 & 0.055 & 8.8E-09 ($mm^2$)\\
\hline
\textbf{Adder/Subtractor \cite{astra}} & 0.0028 & 0.7199 & 5.5E-09 ($mm^2$)\\
\hline
\textbf{Look-up Table (LUT) \cite{astra}} & 1.403 & 0.2225 & 1.597E-06 ($mm^2$) \\
\hline
\textbf{ADC (1 GS/s) \cite{adc1}} & 2.55 & 0.78 & 2E-3 ($mm^2$) \\
\hline
\textbf{ADC (10 GS/s) \cite{adc2}} & 30 & 0.78 & 103E-3 ($mm^2$)\\
\hline
\textbf{DAC \cite{dac}} & 50 & 0.0714 & 11000 ($\mu m^2$)\\
\hline
\textbf{Serializer \cite{lin20250serial}} & 1.5 & 0.03 & 0.0021 ($mm^2$)\\
\hline
\end{tabular}%
}
\end{table}

\section{Results and Comparative Analysis}


We compare the six photonic configurations against two enterprise CPUs (Xeon 6980P, EPYC 9654), an NVIDIA H100 SXM5 GPU \cite{h100gpu}, and a TPU v5p \cite{schneider2025lifecycleemissionsaihardwaretpu}. CPU embodied carbon is estimated with ECO-CHIP \cite{eco}, and GPU and TPU embodied carbon from published GPU and TPU lifecycle footprints.
CPU, GPU, and TPU baselines are estimated using a roofline model with vendor-specified peak throughput, memory bandwidth, and power, on the same transformer workloads at batch size one \cite{williams2009roofline}.

\subsection{Area and Embodied Carbon}
\begin{figure}[htbp]
    \centering
    \includegraphics[width=\columnwidth]{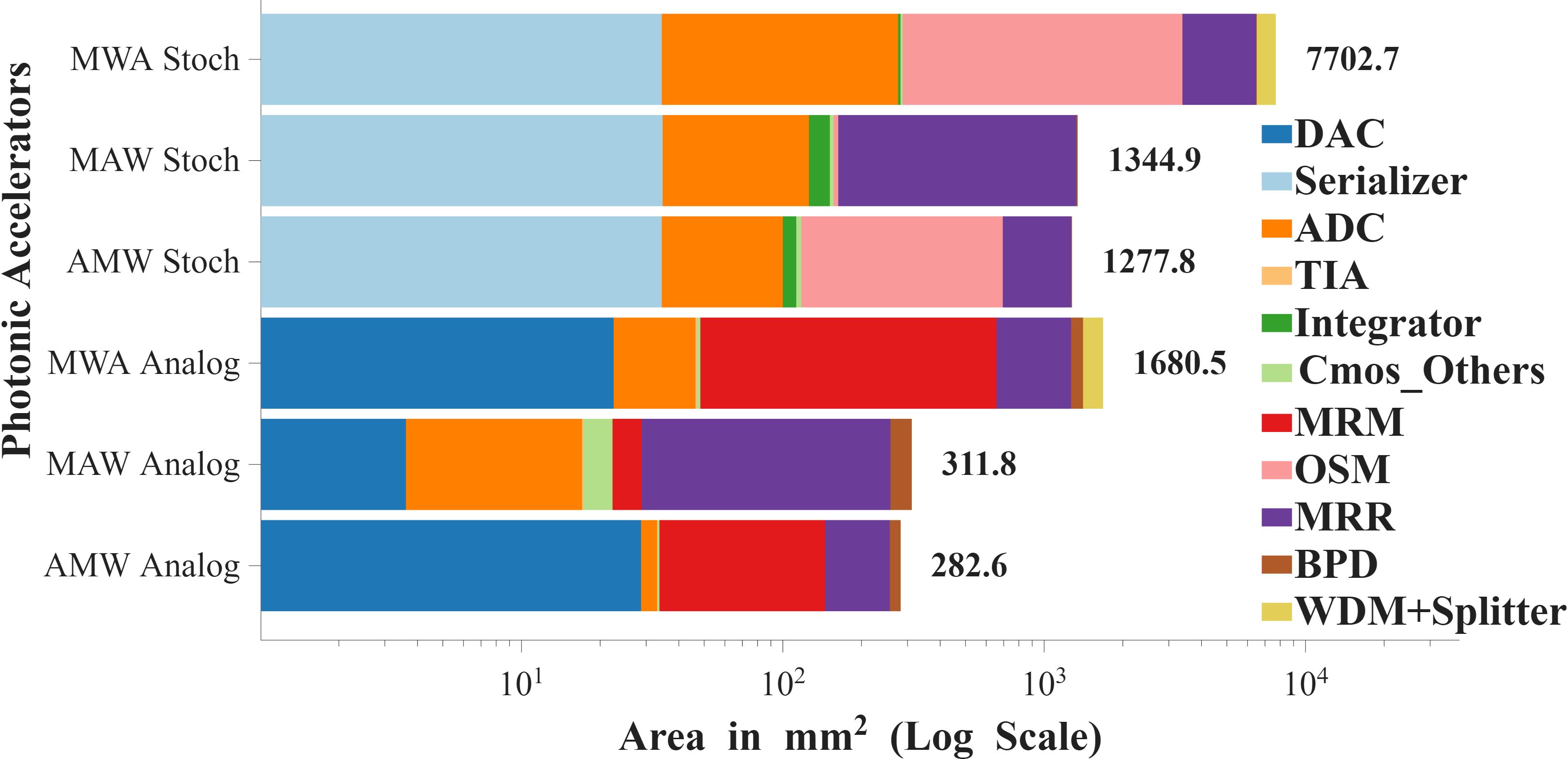}
    \captionsetup{font=small}
    \caption{Component-wise die area for the six configurations.}
    \label{area_stack}
\end{figure}

Stochastic operation trades area for energy: analog-to-stochastic conversion increases die area by 4.3--4.6$\times$, with AMW increasing from 282.6 to 1277.8~$mm^2$, MAW from 311.8 to 1344.9~$mm^2$, and MWA from 1680.5 to 7702.7~$mm^2$. The serializer and, secondarily, the OSM drive this growth because both occupy more area than the DAC and ADC circuitry they replace (Figure~\ref{area_stack}).

\vspace{5pt}

\begin{figure}[htbp]
    \centering
    \includegraphics[width=\columnwidth]{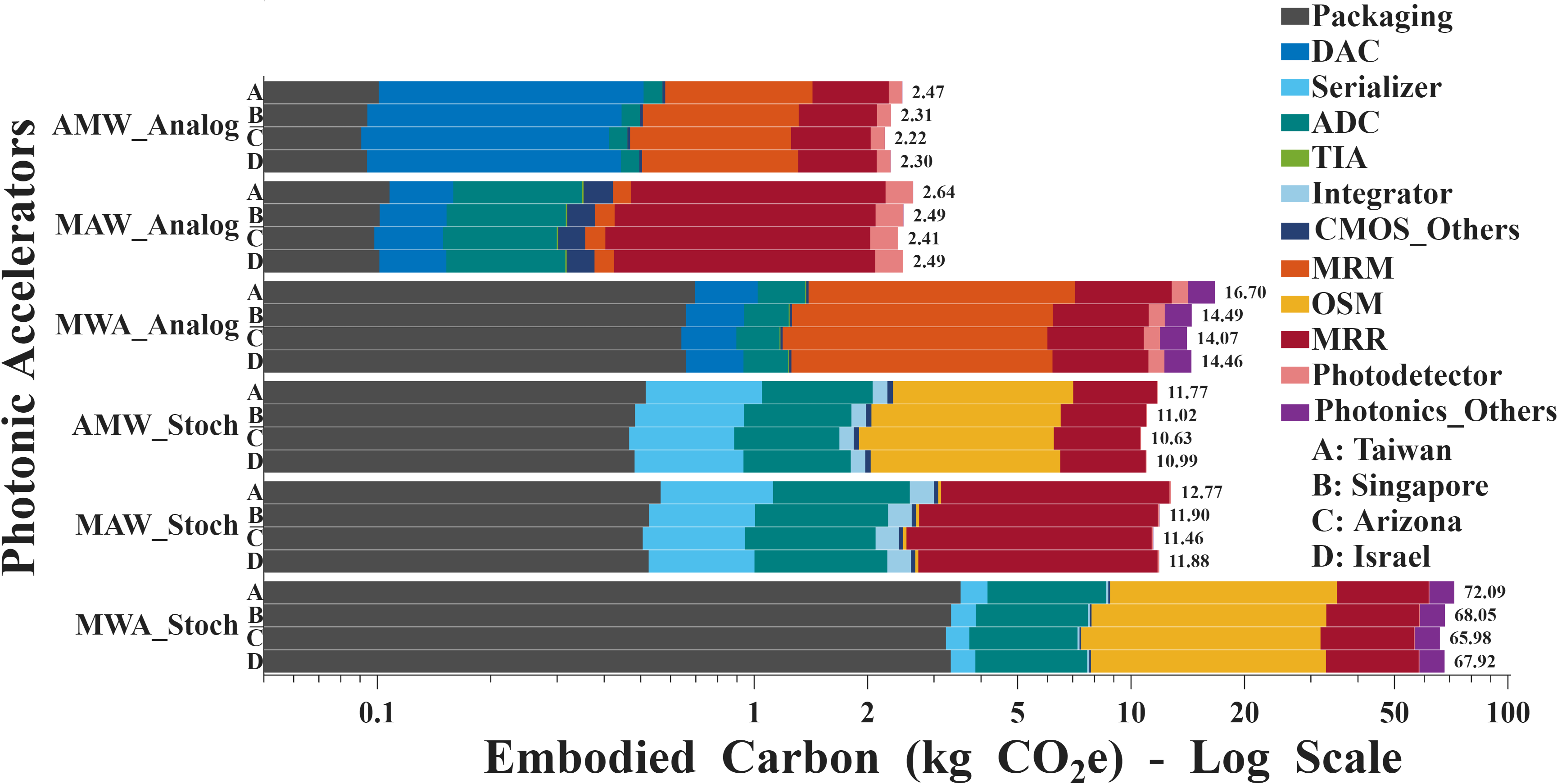}
    \captionsetup{font=small}
    \caption{Component-wise embodied carbon across four foundry grids and the RFCW operational grid.}
    \label{stacked_carbon}
\end{figure}

Embodied carbon increases by 4.4--4.8$\times$ from analog to stochastic across the four foundry grids in Table~\ref{tab:grid_ci}, but varies by only 10--12\% across grids within each configuration (Figure~\ref{stacked_carbon}). Thus, die area dominates fabrication location.

\vspace{5pt}

\vspace{5pt}
\begin{figure}[htbp]
    \centering
    \includegraphics[width=\columnwidth]{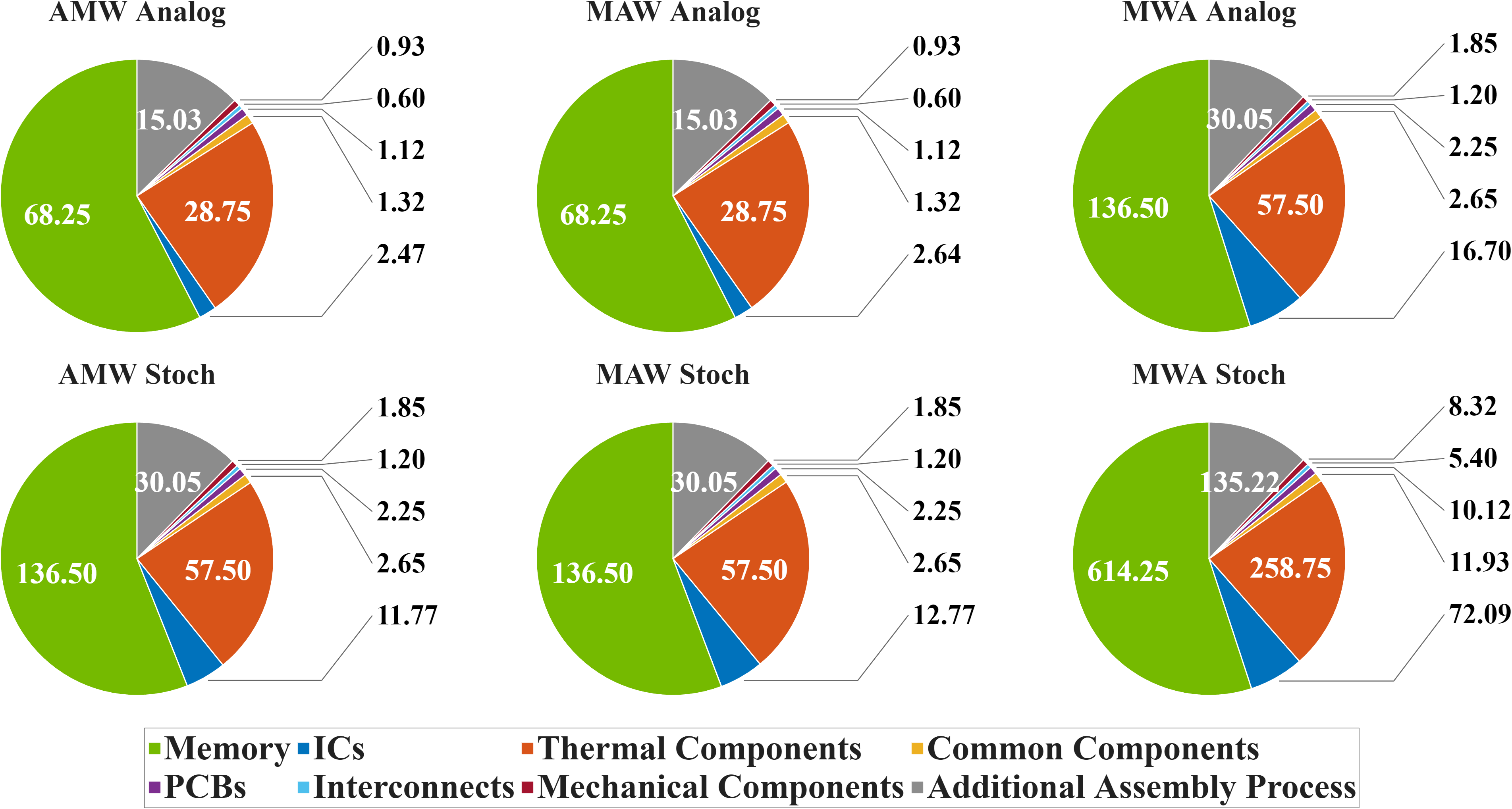}
    \captionsetup{font=small}
    \caption{Deployment-scale embodied carbon the six configurations, derived from \cite{h100gpu} and scaled by photonic chip count.}
    \label{package_pi_chart}
\end{figure}
\vspace{5pt}

At deployment scale, die share falls to a few percent. Lacking a photonic full-system footprint, we estimate deployment-scale embodied carbon from the NVIDIA HGX H100 product carbon footprint (PCF) \cite{h100gpu}: we assign one-eighth of the baseboard PCF, replace the H100 SXM5 die with our photonic chip, and scale memory, thermal, and assembly by chip count. Memory dominates, followed by thermal components and assembly (Figure~\ref{package_pi_chart}). Die share spans 2.1\% (AMW-Analog, 2.47~kg CO$_2$e) to 6.7\% (MWA-Analog, 16.70~kg CO$_2$e), versus 6.5\% for MWA-Stoch. Totals span 118.5~kg CO$_2$e (AMW-Analog and MAW-Analog) to 1116~kg CO$_2$e (MWA-Stoch), a chip-count-driven spread of 9.4$\times$.

\textit{Takeaway: organization and signaling mode determine die area and embodied carbon; fabrication location is secondary.}

\begin{figure*}[!ht]
    \centering
    \begin{subfigure}[b]{0.32\textwidth}
        \includegraphics[width=\textwidth]{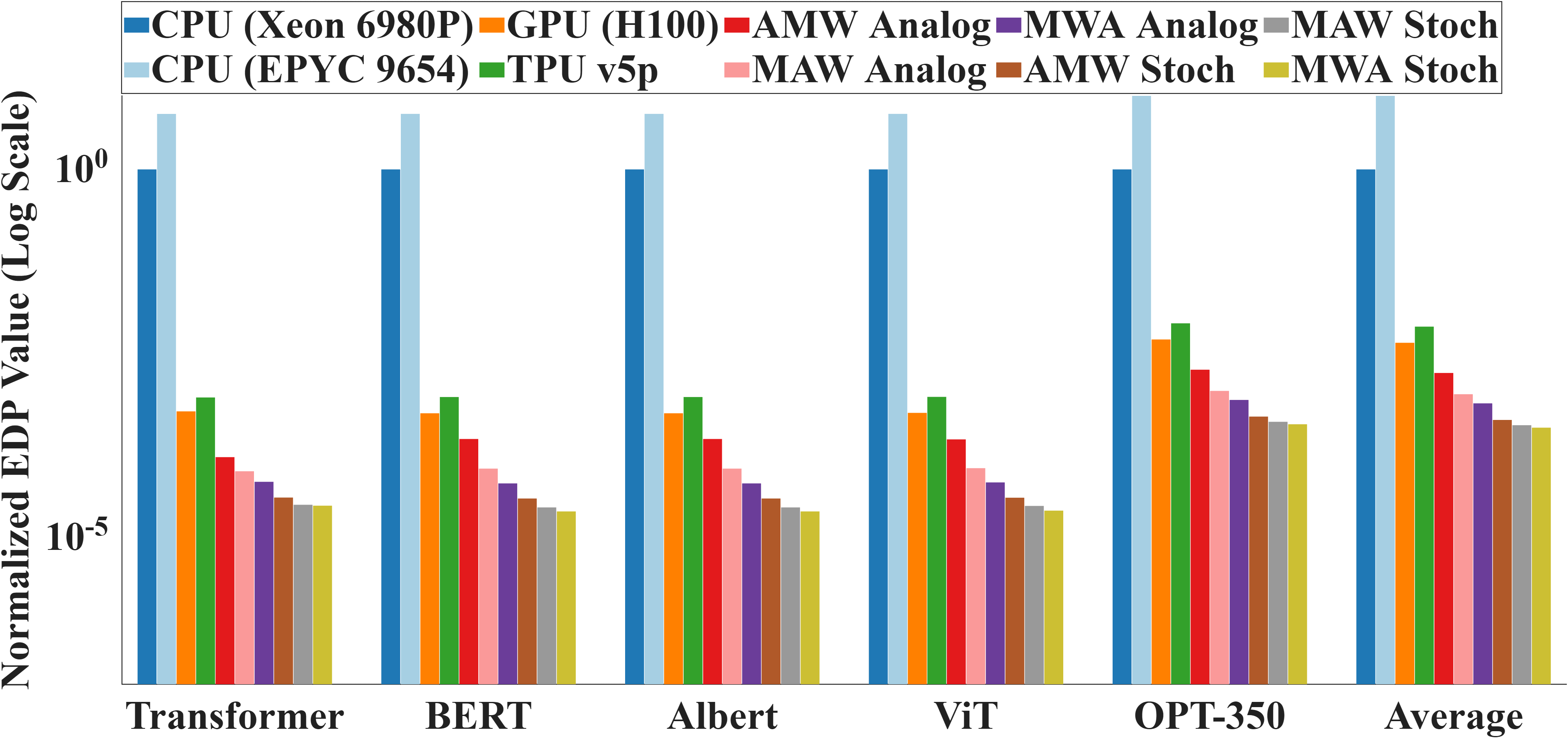}
        \caption{EDP}
        \label{edp}
    \end{subfigure}
    \hfill
    \begin{subfigure}[b]{0.32\textwidth}
        \includegraphics[width=\textwidth]{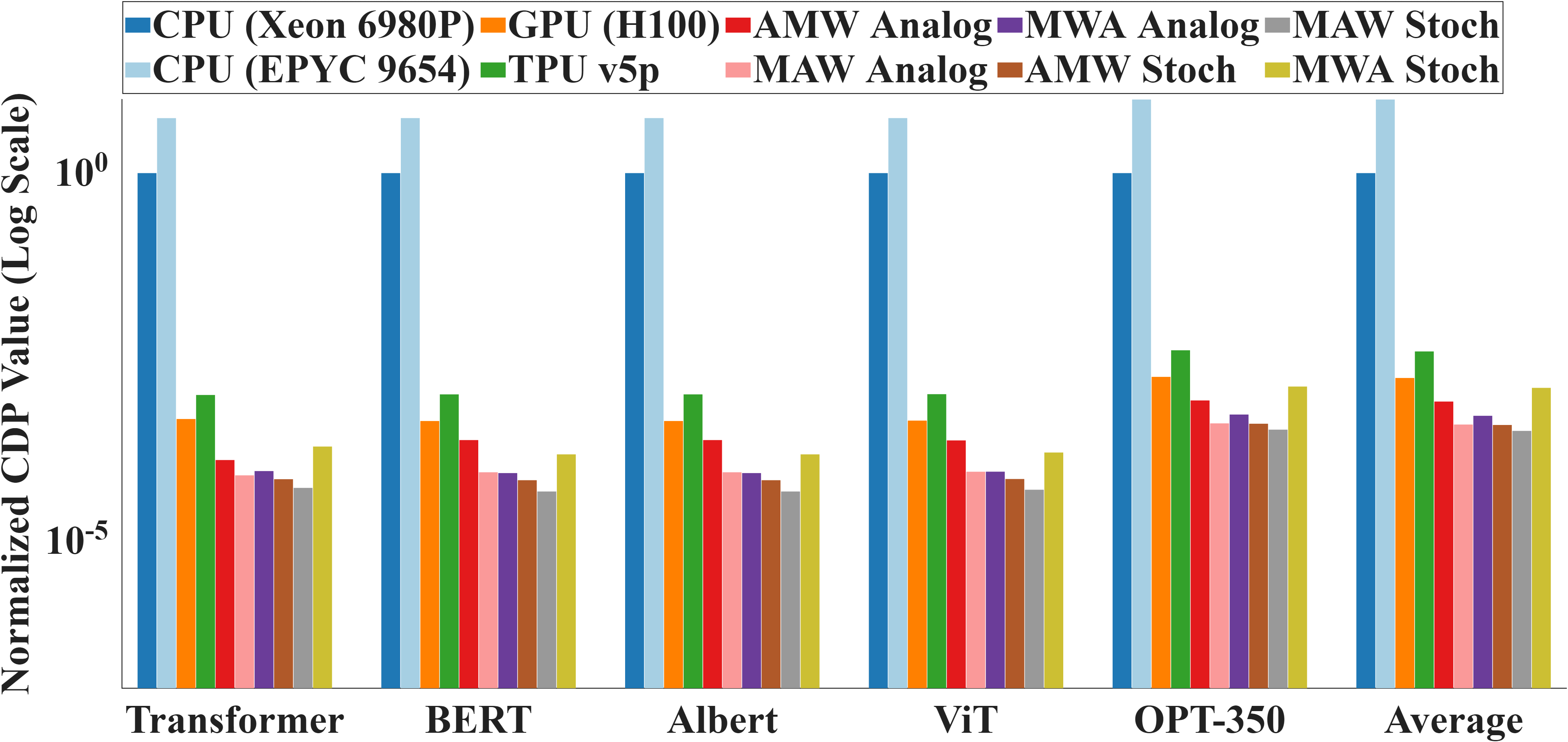}
        \caption{CDP}
        \label{cdp}
    \end{subfigure}
    \hfill
    \begin{subfigure}[b]{0.32\textwidth}
        \includegraphics[width=\textwidth]{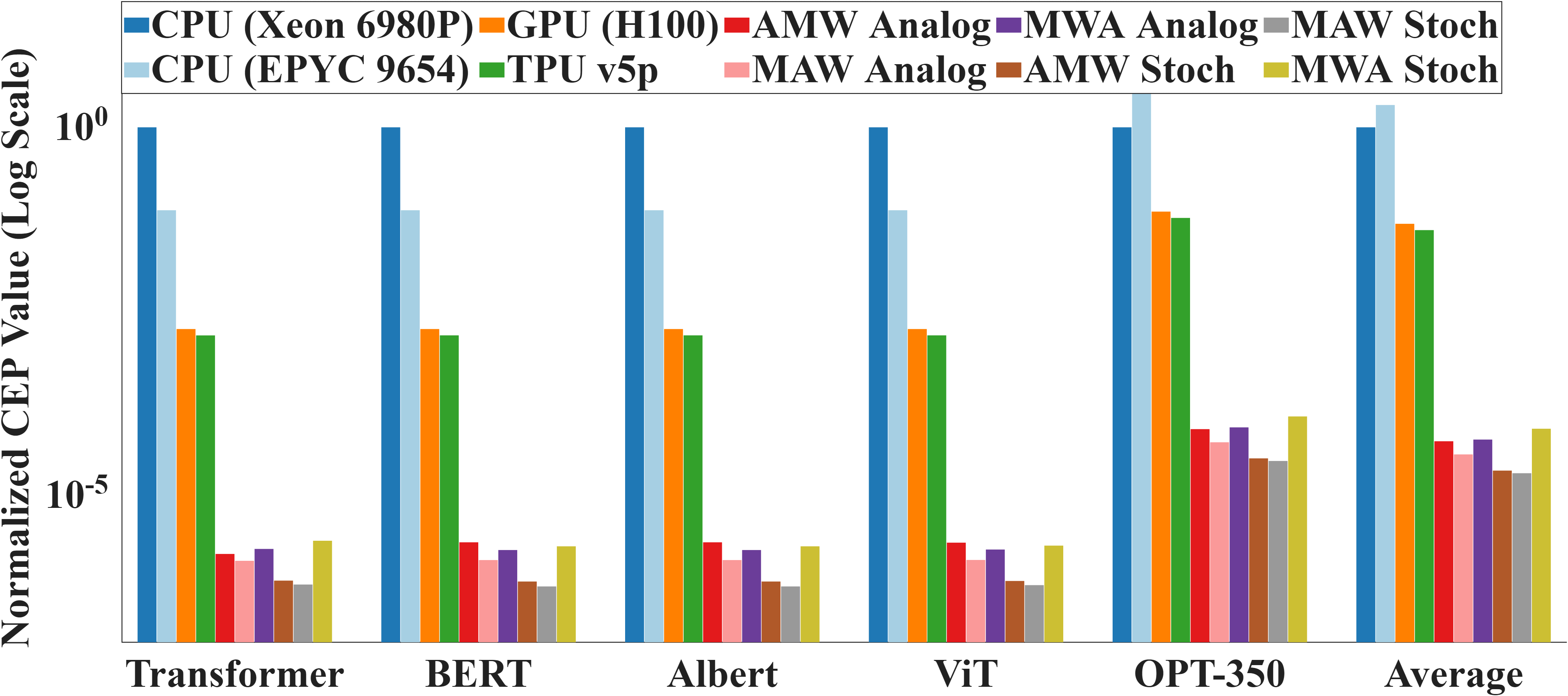}
        \caption{CEP}
        \label{cep}
    \end{subfigure}
    \captionsetup{font=small}
    \caption{Normalized (a) EDP, (b) CDP, and (c) CEP for ten platforms and five models (log scale, normalized to Xeon 6980P). Photonic designs reach four orders of magnitude lower EDP and CDP and six orders lower CEP than the CPUs.}
    \label{fig:metrics}
\end{figure*}

\vspace{5pt}
\begin{figure}[htbp]
    \centering
    \includegraphics[width=0.9\columnwidth]{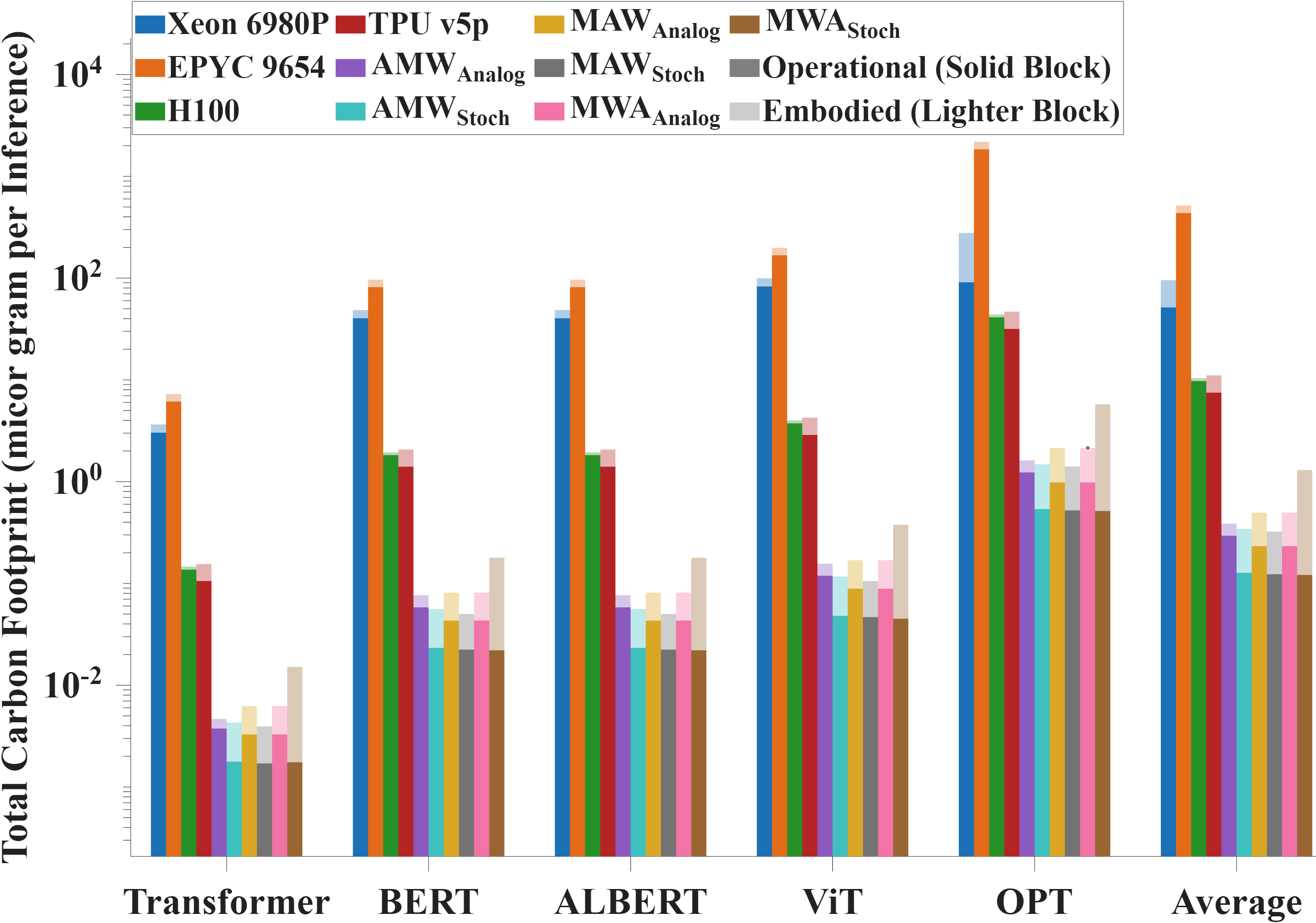}
    \captionsetup{font=small}
    \caption{Per-inference carbon ($\mu$g CO$_2$e) for ten platforms and five models (log scale), split into operational (solid) and embodied (light).}
    \label{cfp}
\end{figure}

\subsection{EDP, CDP, and CEP}
Carbon-aware metrics reorder the performance ranking (Figure~\ref{fig:metrics}, normalized to Xeon 6980P). On EDP, which is carbon-blind, photonic architectures are about four orders of magnitude below the CPUs and two below the H100 SXM5 and TPU v5p; within the photonic cluster, stochastic beats analog, and MWA-Stoch is lowest across all five models. Under CDP the ranking flips: MAW-Stoch is lowest overall, while MWA-Stoch becomes the worst stochastic design, its larger embodied carbon offsetting its latency advantage. Under CEP, photonic architectures reach six orders of magnitude below the CPUs and four below the GPU and TPU on the four smaller models, with MAW-Stoch again lowest. For all three metrics, the move to optical processing preserves photonic designs’ advantage over conventional counterparts at the large model scale defined by OPT-350. Since larger batches, models, and sequence lengths scale the workload equally across all six configurations, the relative comparison at batch size one is expected to hold, although growing memory and weight-reprogramming overheads may narrow the gains.

\textit{Takeaway: the performance-optimal (MWA-Stoch) and carbon-optimal (MAW-Stoch) designs differ; MAW-Stoch is the balanced choice for sustainable edge inference.}

\subsection{Per-Inference Carbon}
Combining embodied carbon with operational energy yields the per-inference footprint that matters at the edge, and at this granularity the operational-embodied balance inverts between platform classes. For CPU, GPU, and TPU baselines, operational carbon dominates, exceeding embodied by more than three orders of magnitude for CPUs and two for the H100 SXM5 and TPU v5p. For every photonic configuration, embodied carbon instead matches or dominates operational carbon across all five models, because photonic operational energy is low enough that amortized manufacturing carbon dominates it (Figure~\ref{cfp}). MAW-Stoch achieves the lowest total per-inference carbon of any platform, followed by MAW-Analog and AMW- Stoch, by balancing low die area with stochastic-mode energy efficiency. 

\textit{Takeaway: embodied carbon is a first-order term for edge photonic inference, and MAW-Stoch minimizes total per-inference carbon.}

\section{Conclusion}
 This work establishes photonic transformer accelerators as viable for edge deployment and shows that design selection must consider more than speed and energy. We evaluated six MRR-based DPU configurations (AMW, MAW, MWA in analog and stochastic modes) against CPU, GPU, and TPU baselines on five transformer models, measuring per-inference operational and embodied carbon with 2.5D interposer overheads across four foundries. \textit{Three} insights emerge. \textit{First}, photonic designs achieve up to \(10^4\times\) lower EDP and CDP and \(10^6\times\) lower CEP than CPUs.
\textit{Second}, analog-to-stochastic conversion increases area and embodied carbon by \(4.5\times\) while reducing operational energy by 91\%, often causing embodied carbon to exceed operational carbon per inference. \textit{Third}, fastest is not always most sustainable: MWA-Stoch minimizes EDP but has the highest embodied carbon while MAW-Stoch minimizes CDP, CEP, and total per-inference carbon. Organization and signaling rather than location dictates this outcome, enabling carbon-aware designs that simultaneously account for manufacturing and operational efficiency.



\section*{Acknowledgments}
The authors used Claude (Anthropic) to improve the clarity and grammatical syntax in selected sections of this manuscript. All technical content, results, figures, analyses, and conclusions are the sole work of the authors.

\bibliographystyle{IEEEtran}
\bibliography{references}

\end{document}